# On-chip broadband plasmon-induced transparency based on plasmonic metal-insulator-metal waveguides

Zhaojian Zhang [1], Junbo Yang [2, *], Siyu Xu [1], Heng Xu [1], Yunxin Han [2], Xin He [2], Jingjing Zhang [1], Jie Huang [1] and Dingbo Chen [1]

[1] College of Liberal Arts and Sciences, National University of Defense Technology, Changsha 410073, China; 376824388@qq.com

[2] Center of Material Science, National University of Defense Technology, Changsha 410073, China; yangjunbo@nudt.edu.cn

* Correspondence: yangjunbo@nudt.edu.cn

**Abstract**
As an analogue of electromagnetically induced transparency (EIT), plasmon-induced transparency (PIT) has been realized both in plasmonic metamaterial and waveguide structures. Via near-field coupling within unit cells, PIT with broadband could be produced by plasmonic metamaterials, which, however, has not been realized in on-chip plasmonic waveguide structures. Here, we introduce broadband PIT based on a plasmonic metal-insulator-metal (MIM) waveguide system. Utilizing the direct coupling structure, PIT emerges based on an easy-fabricated structure without gap. By tuning coupling distance, the transparent window can be continuously tuned from narrow- to broadband. Such device is promising for on-chip applications on sensing, filtering and slow light over a broad frequency range.

## 1. Introduction

Electromagnetically-induced transparency (EIT) is a quantum phenomenon arising from the interference within two different excitation channels in a laser-driven atomic system [1], and has been focused due to its applications on slow light [2], nonlinear enhancement [3] and optical storage [4]. However, EIT is hampered by cumbersome experimental conditions such as stable pumping and low temperature [1]. Recently, by mimicking the generation process of EIT, a plasmonic system can easily produce EIT-like effect via the interaction between the bright and dark modes [5], which has been called plasmon-induced transparency (PIT). For now, PIT has been achieved based on plasmonic metamaterials [5] and waveguide systems [6], paving the way to realize various photonic applications such as sensors [7-8], filters [9-10], switches [11-12] and slow light devices [13-14].

However, the classical PIT always possesses a narrow transparent window, which limits its applications over a broadband range. To overcome such obstacle, multi-PIT, which has multiple narrow transparent windows, has been introduced into plasmonic metamaterial and waveguide systems via the coupling of multiple electromagnetic modes [15-16]. Furthermore, utilizing the constructive interference between electric and magnetic dipole modes, broadband PIT has also been realized based on metamaterials [17-18]. Such broadband PIT has a transparent window across a wide spectral range, promising for broadband photonic devices. However, few works have been reported to realize broadband PIT in plasmonic waveguide structures to meet the requirement of applications at chip scale. Recently, as a promising candidate for achieving on-chip all-optical circuits in the future, metal-insulator-metal (MIM) waveguide has been extensively studied [6, 8,

10, 12, 14, 19-20]. MIM waveguide can confine the light into subwavelength scale, simultaneously possessing easy fabrication.

In the work, we numerically propose the broadband PIT based on a MIM waveguide system. At first, a novel direct coupling structure is introduced to realize the classical PIT, the corresponding theoretical description is given by coupled mode theory (CMT). And then, via tuning the coupling distance of two resonators, the transparent window can be continuously tuned from narrow- to broadband. Finally, the performance of slow light is studied based on broadband PIT. This work introduces broadband PIT into the plasmonic waveguide structures, and has potential on-chip applications on broadband sensing, filtering and slow light.

## 2. Models and theories

The structures are presented in the form of 2D as shown in Fig. 1 (a-b), the grey area is silver (Ag), and white area represents air. The geometric parameters are given in the caption of Fig. 1. The dielectric constant of Ag is calculated by the Drude model [21]:

$$\varepsilon_m = \varepsilon_\infty - \frac{\omega_p^2}{\omega(\omega+i\gamma)} \quad (1)$$

Where $\varepsilon_\infty$ is the dielectric constant for the infinite frequency, $\omega_p$ refers to the bulk frequency of plasma, $\gamma$ is the damping frequency of electron oscillation, and $\omega$ gives angular frequency for the input light wave. For Ag, these parameters are $\varepsilon_\infty = 3.7$, $\omega_p = 1.38 \times 10^{16}$ Hz, and $\gamma = 2.73 \times 10^{13}$ Hz.

In the MIM waveguide structure, there is only transverse-magnetic (TM) mode [22]. The width of the waveguide is much smaller than the incident wavelength, consequently only fundamental TM mode exists. The dispersion of such fundamental mode is derived as follows [22]:

$$\frac{\varepsilon_i p}{\varepsilon_m k} = \frac{1-e^{kw}}{1+e^{kw}}$$

$$k = k_0 \sqrt{(\frac{\beta_{spp}}{k_0})^2 - \varepsilon_i}, p = k_0 \sqrt{(\frac{\beta_{spp}}{k_0})^2 - \varepsilon_m} \quad (2)$$

$$\beta_{spp} = n_{eff} k_0 = n_{eff} \frac{2\pi}{\lambda}$$

Where $w$ is the waveguide width, $\lambda$ is vacuum wavelength of the input light, and $\varepsilon_i$ and $\varepsilon_m$ is dielectric and metal permittivity, respectively. $\beta_{spp}$ is the propagation constant of surface plasmon waves, $n_{eff}$ represents effective refractive index inside the waveguide, and $k_0 = 2\pi/\lambda$ is wavenumbers. 2D Finite-Difference Time-Domain (FDTD) method is used to calculate the optical response of proposed structures, which will have the same results as 3D simulation when the height (along the z direction) of the structure is more than 1 um [19]. This is because 2D FDTD simulation considers the third direction (z direction) as infinite in physics [20]. the mesh size is 2.5 nm, and the boundary condition is set as perfectly matched layers (PML) to ensure convergence. As given in Fig. 1(a-b), The light source is set at the position of $P_{in}$, and one monitor is set at $P_{out}$ to gather power and phase data of output light. The transmission spectrum of power is defined as $T = P_{out}/P_{in}$.

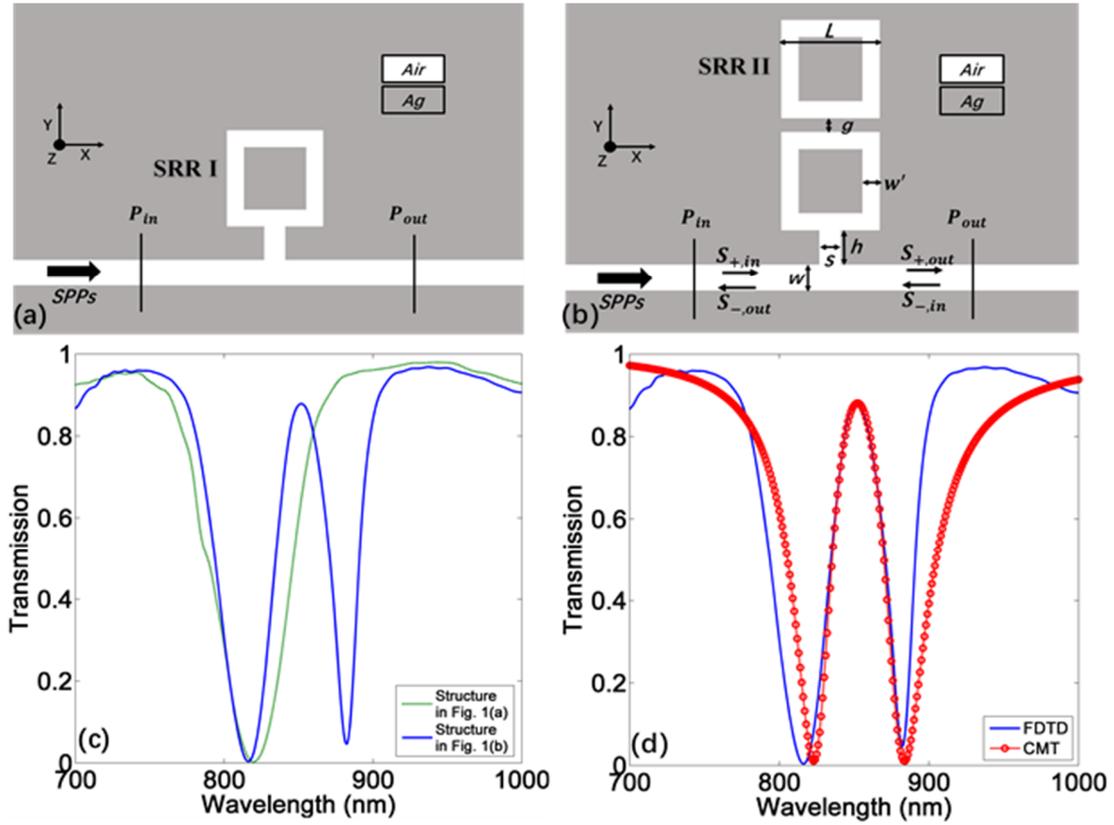

Fig. 1 (a) The bus waveguide is aperture-coupled with single SRR. (b) The second SRR is indirectly coupled at the upper side of the first SRR. Structural parameters: *w*= 100 nm, *s*= 80 nm, *h*= 130 nm, *w'*= 50 nm, *L*= 375 nm, the gap width is *g*= 15 nm. (c) The transmission spectrum corresponding to the two structures. (d) The transmission spectrum from FDTD and CMT.

Fig. 1(a) shows a bus waveguide side-coupled with s square ring resonator (SRR, here is SRR I) via an aperture. The resonant wavelength of SRR I is 820 nm, which is indicated in Fig. 1(c) by the transmission spectrum in green, and such resonant mode is a typical whispering gallery mode (WGM). The resonance condition is derived as follows [12]:

$$m\lambda = 4L_{eff} \operatorname{Re}(n'_{eff}), m = 1, 2... \quad (3)$$

Where $n'_{eff}$ is effective refractive index of the waveguide, which can be solved by Eq. (2). $L_{eff}$ is effective side length of SRR, which is the average of the inner and outer side length. *m* is the integer mode number. When another same SRR, SRR II, is indirectly coupled above SRR I as shown in Fig. 1(b), the transmission spectrum in Fig. 1(c) will turn to a classical PIT profile, which arises from the interference [23]. The mode in SRR I is directly excited by light source and acts as the bright mode, and the mode in SRR II, which is excited by SRR I, is the dark mode indirectly excited by source. The interaction between bright and dark modes will bring about the coherent interference between the direct and indirect excitation pathways, leading to a transparent window in the transmission spectrum. Such process is similar with the quantum interference in a three-level atomic system producing EIT [1]. However, PIT can also be described equivalently as two supermodes with close resonant wavelengths caused by the coupling between two basic modes [24].

In MIM waveguide system, PIT can also be theoretically analyzed by CMT [25]. As presented in Fig. 1(b), $S_{\pm,in}$ and $S_{\pm,out}$ are amplitudes of input and output waves respectively, the subscript $\pm$ are opposite directions of propagation. The decay rates of two SRRs, caused by intrinsic loss, are set as $\alpha$ and $\beta$ respectively. The coupling coefficient between the bus waveguide and SRR I is described as $\gamma$, $\delta$ refers to the coupling coefficient between two SRRs. According to temporal CMT, the field amplitudes, *a* for bright mode and *b* for dark mode, can be calculated as [25]:

$$\frac{da}{dt} = (j\omega_0 - \alpha - \gamma)a + j\sqrt{\gamma}(S_{+in} + S_{-in}) + j\delta b$$
$$\frac{db}{dt} = (j\omega_0 - \beta)b + j\delta a$$
(4)

Here, $\omega_0$ is the common resonant frequency of two SRRs, and *j* is the imaginary unit. According to power conservation and time reversal symmetry, the input and output wave amplitudes have a relation:

$$S_{+,out} = S_{+,in} + j\sqrt{\gamma}a$$
$$S_{-,out} = S_{-,in} + j\sqrt{\gamma}a$$
(5)

Where $S_{-,in}= 0$ because light is input from the left. Based on Eq. (4-5), the transmission spectrum can be given as:

$$T(\omega) = \left|\frac{S_{+,out}}{S_{+,in}}\right|^2 = \left|1 - 2\gamma\frac{2j(\omega-\omega_0)+2\beta}{[2j(\omega-\omega_0)+\alpha+\beta+\gamma]^2+(2\delta)^2-(\alpha-\beta+\gamma)^2}\right|^2$$
(6)

The transmission spectra for the structure in Fig. 1(b) are plotted in Fig. 1(d) from both FDTD and CMT, indicating a good fitting between theoretical and simulation results.

It has to be mentioned that the design scheme of Fig. 1(b) is conventional for PIT implementation in MIM waveguides, which always includes indirectly coupled resonators with different shapes and are based on 2D FDTD simulation [6, 8, 10, 12, 14, 23-25]. However, such indirect coupling structure will possess one or several metal gaps with a dozen nanometers in width (defined as *g* in Fig. 1(b)). As we have mentioned, the 2D simulation result is equal to that in 3D simulation only if the height of structure is more than 1 um. Consequently, the metal gap will become an ultra-thin metal wall in 3D structure, which are hardly fabricated in practice. Therefore, most similar works are only based on simulation.

**3. PIT in direct coupling structure**

Here, we introduce a novel direct coupling structure as shown in Fig. 2(a). The width of gap is reduced to zero, leading a direct coupling area with distance *d*= 100 nm. The corresponding transmission spectrum in Fig. 2(c) indicates that PIT still happens in such structure and has a blueshift compared with PIT in indirect coupling structure. The $H_z$ distributions at dip points of PIT both in direct coupling and indirect coupling structures are shown in Fig. 2(b) for comparison. In the indirect coupling structure, the WGMs in two SRRs has in-phase and out-phase distributions at c and d points respectively, which can be regarded as two different supermodes. When the gap is removed, two supermodes still exist with a slight blueshift that comes from the enhancement of coupling strength between two WGMs, which is called coupling-induced resonance frequency shifts (CIFS) [26]. The supermode distribution at the left dip is still in-phase, another supermode, however, becomes different. Such phenomenon will be discussed in the next part. Notably, this structure possesses easy-fabrication and low-loss simultaneously due to a lack of metal gaps, thus is valuable for realizing on-chip PIT in practice.

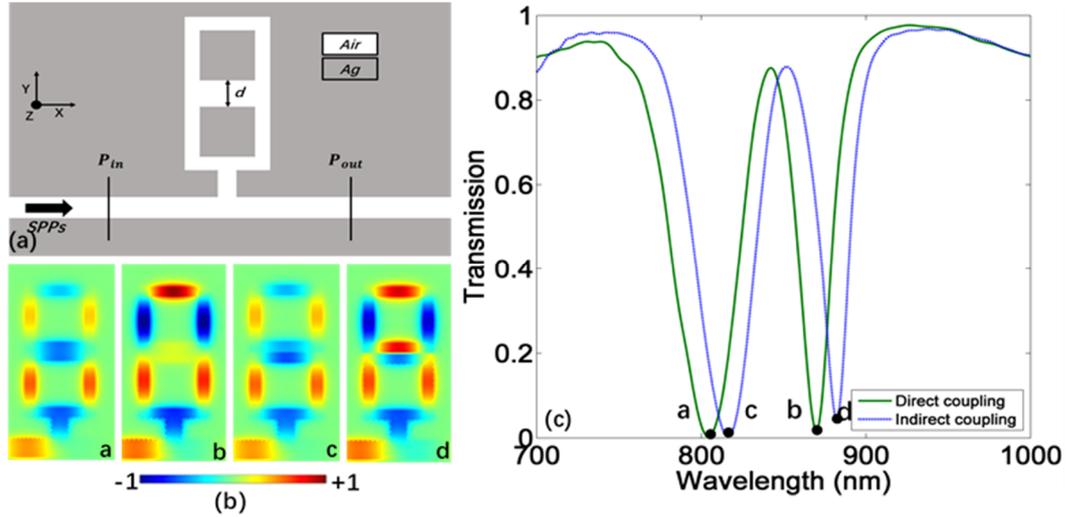

Fig. 2 (a) The structure with direct coupling structure, $d$= 100 nm. (b) The $H_z$ distributions for wavelength points in (c). (c) The transmission spectrum corresponding to direct and indirect coupling structures.

Furthermore, we find that if the coupling distance is increased, the transparent peak will become wider. The bandwidth of PIT is evaluated by the full wave at half maximum (FWHM) of the transparent window. In the original structure, FWHM of the transparent peak is 35.1 nm with $d$= 100 nm. However, when $d$ comes to 200 nm, FWHM will be increased to 55.1 nm as shown in Fig. 3 (a). As $d$ increasing, the transparent window will be continuously broadened as depicted in Fig. (b-c), the relation between FWHM and $d$ is shown in Fig. 4, which indicates a nonlinear relationship. Thus, an on-chip broadband PIT is realized based on such direct coupling structure, and FWHM of transparent window can be easily tuned by changing the coupling distance.

The $H_z$ distributions at two transmission dips (two supermodes) in structures with different coupling distances are also given in Fig. 3. From the magnetic fields at a, c and e points we can see that the WGMs still exist in SRRs and construct the left supermode, despite the change of coupling area. This is because that the whole coupling area has a magnetic field with a negative direction, which can help construct the WGM in SRR. However, the magnetic fields at b, d and f points shows that there is not WGM inside SRRs. Especially, from the magnetic field at f point we can see that the whole resonator supports a special standing-wave mode (SWM) rather a supermode, alternatively.

We also notice that the supermode at the left dip in the spectrum has a slighter redshift compared with the right mode when coupling distance is increased, which lead to a broader transparent peak. Basically, the left supermode is a WGM-based supermode, the resonance condition of WGM is related to the side length of SRR. If the coupling area is regarded as the fourth side of SRR, only the horizontal width of the coupling area is related to the fourth-side length, so an increasing of vertical length (coupling distance) only weakens the coupling strength between two WGMs. Therefore, the slight redshift of left supermode is induced by the change of coupling strength. However, for the SWM at the right dip, the resonance condition is proportional and sensitive to the vertical length of coupling area, since such mode is formed by the back-and-forth reflection along the vertical direction. Therefore, SWM has a more obvious redshift when coupling distance is increased.

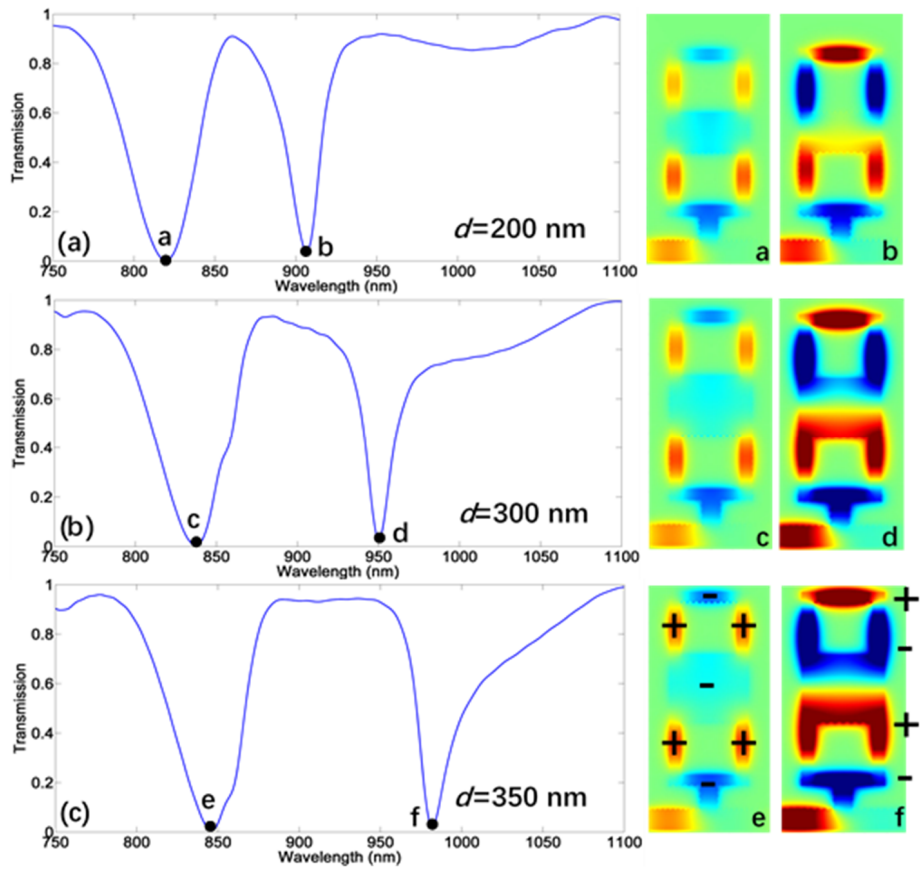

Fig. 3 (a) The transmission spectrum when $d=$ 200 nm and corresponding $H_z$ distributions at two dips. (b) The transmission spectrum when $d=$ 300 nm and corresponding $H_z$ distributions at two dips. (c) The transmission spectrum when $d=$ 350 nm and corresponding $H_z$ distributions at two dips.

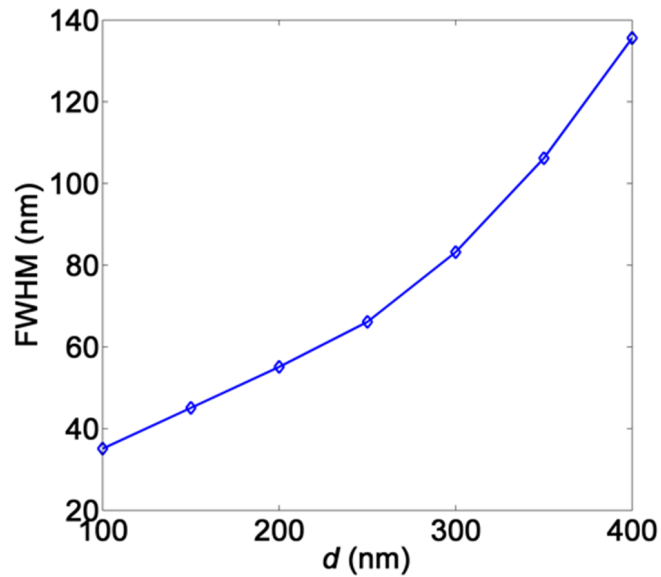

Fig. 4 The relation between the FWHM and coupling distance $d$.

## 4. Broadband slow light

One of the most conventional applications of PIT is slow light. A sharp transmission peak always indicates that there exists intense dispersion in the transparent window, which will cause the group velocity delay [16]. The performance of slow light is assessed by the group index $n_g$ calculated as follows [16]:

$$\tau_g = \frac{d\psi(\omega)}{d\omega}$$
$$n_g = \frac{c}{v_g} = \frac{c}{D}\tau_g \qquad (7)$$

Here, $\psi(\omega)$ is the transmission phase shift from the light source to the monitor, $c$ stands for the light speed, $v_g$ refers to the group velocity in the waveguide, and $D$ is the length of this system, which is 500 nm. Fig. 5(a-b) indicate the phase shift and group index of this structure under different coupling distances, and we can find that a wider transparent window will provide a broader range of slow light. Meanwhile, however, the group index will decrease. This is because that a broader transmission peak will lead to a weaker dispersion, thus reducing the delay time of slow light. Therefore, there is a trade-off between the strength and range of slow light.

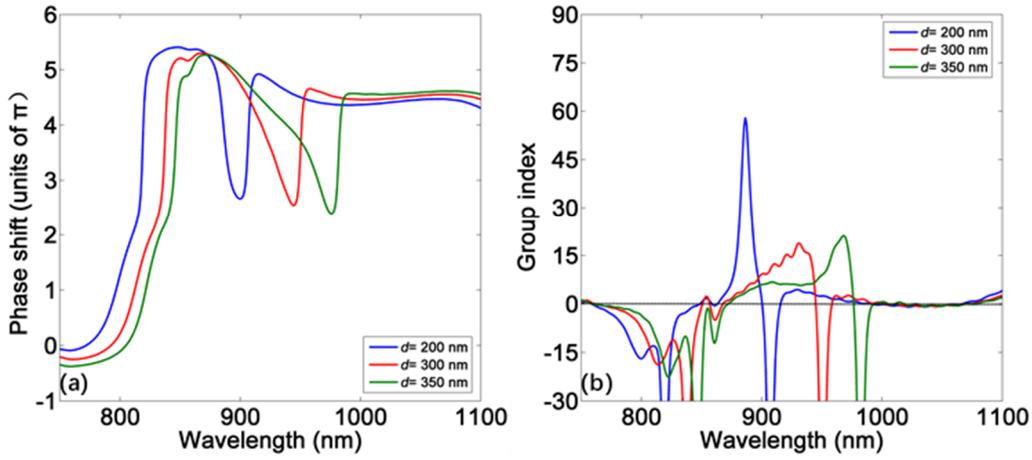

Fig. 5 (a) The phase shift with different coupling distances. (b) The group index with different coupling distances.

## 5. Conclusion

In summary, on-chip broadband PIT has been realized based on MIM waveguides with a no metal-gap structure. Such structure not only possesses low intrinsic loss, but also can greatly simplify the fabrication, which is a valuable approach for experiments. More importantly, broadband PIT has been introduced into a MIM waveguide system, which is promising for on-chip broadband applications. This work has potential applications on sensing, filtering, nonlinear enhancement and slow light over a broad frequency range in the future highly integrated all-optical circuits.

**Acknowledgments**

This work is supported by the National Natural Science Foundation of China (61671455, 61805278), the Foundation of NUDT (ZK17-03-01), the Program for New Century Excellent Talents in University (NCET-12-0142), and the China Postdoctoral Science Foundation (2018M633704).